\documentclass[prd, tightenlines, twocolumn]{revtex4-1}
\usepackage{amsmath}
\usepackage{tipa}
\usepackage{amssymb}
\usepackage{graphicx}
\usepackage{xfrac}
\usepackage{bbm}
\usepackage{enumerate}
\usepackage{color}
\usepackage[colorlinks = true,
            linkcolor = blue,
            urlcolor  = blue,
            citecolor = blue,
            anchorcolor = blue]{hyperref}
\usepackage{subfigure}
\usepackage{physics}
\usepackage{rotating}
\usepackage{capt-of}
\usepackage{chngcntr}
\usepackage{varwidth}
\usepackage{slashed}

\newcommand{\be}{\begin{equation}}
\newcommand{\ee}{\end{equation}}
\newcommand{\bea}{\begin{eqnarray}}
\newcommand{\eea}{\end{eqnarray}}
\newcommand{\ba}{\begin{eqnarray}}
\newcommand{\ea}{\end{eqnarray}}

\begin{document}

\title{Extending the hydrodynamical description of heavy-ion collisions\\
 to the  ``outer edge" of the fireball}
\author{ Adith Ramamurti}
\email{adith.ramamurti@stonybrook.edu}
\author{ Edward Shuryak }
\email{edward.shuryak@stonybrook.edu}

\affiliation{Department of Physics and Astronomy, \\ Stony Brook University,\\
Stony Brook, NY 11794, USA}

\date{\today}

\begin{abstract} 
It is well known that relativistic hydrodynamics provides very good description
of heavy-ion collisions at RHIC/LHC energies up to transverse momenta $p_\perp\sim 2\, \text{GeV}$. 
In this paper, we suggest that this description can be extended to higher $p_\perp\sim 6\, \text{GeV}$,
beyond which hard collisions contribute. While most previous work focused on a 
part of the freezeout surface at the latest proper time (referred to in this work as the ``lid"), we focus
on the complementary part, to be referred as ``the outer edge," where the highest
transverse rapidity of flow $\kappa\sim 1.4$ is achieved. We study this surface analytically, 
using the Riemann rarefaction wave, and numerically, using MUSIC numerical hydrodynamic code. We also  
use an improved freezeout condition, where the collision rate equals the expansion rate. 
For central collisions, we observe good description of spectra for $\pi,K,N$ in central 
PbPb LHC collisions in this extended region.  We further suggest that ``the outer edge" 
has very small azimuthal asymmetry even for non-central collisions, smaller than
predicted by standard hydrodynamics.
 \end{abstract}

\maketitle
\section{Introduction}
It is by now established that high energy heavy-ion collisions, and even 
high-multiplicity $pA$ and $pp$ collisions,
 can be rather
well described by relativistic hydrodynamics; for reviews see Refs. 
\cite{Heinz:2004qz,Shuryak:2014zxa,Weller:2017tsr,Bozek:2013uha}.  These hydrodynamical descriptions of spectra  are typically  accepted from small transverse momenta up to $p_\perp \sim 2\, \text{GeV}$. The reasons given for this upper limit
  differ from paper to paper, and can be summarized as follows:
  \begin{enumerate}[(i)]
  \item the afterburner cascade runs out of statistics;
  \item  the viscous corrections to thermal distribution in the hydro cell may become large \cite{Teaney:2013gca};
   \item  above  $p_\perp> 2-3\, \text{GeV}$  the 
  azimuthal  harmonic flows $v_n(p_\perp)$  no longer follow
  the characteristic hydro-based linear regime  $ v_n(p_\perp)\sim    p_\perp $.

\end{enumerate}
  
%  
%  
%  Later studied has revealed that such growth continues till a maximum, at  $p_\perp^{max}=3-4\, GeV$, after which  $v_n(p_\perp)$ rather rapidly decreases,
%  till it stabilizes at some smaller values, at large $p_\perp> p_\perp^{jet}\sim 10\, GeV$ related with azimuthal
%  dependence of jet quenching.
%
As we will show, the spectra change shape and produce evidence for hard collisions and jet-related phenomena several orders lower, at $p_\perp^\text{max}\sim 5-6\, \text{GeV}$. At present, there exists no generally agreed explanation for the origin of secondaries in the intermediate region of transverse momenta  
  $ 2\, \text{GeV} < p_\perp < p_\perp^\text{max}$. 

The only previous attempt to explain spectra in this region, by a coalescence of jet-related and hydro-related
quarks \cite{Fries:2003vb}, predicted a certain ``quark scaling" of elliptic flow.
Not going into ints criticism, let us just note that our proposal is completely different. 
In particular, we discuss secondaries coming from the freezeout surface at  proper time
$\tau\sim 20\, \text{fm/c}$, whereas all jets leave  the fireball much earlier, at $\tau<6 \, \text{fm/c}$,
so no coalescence is possible.   

Standard application of  -- by now, rather sophisticated -- hydrodynamics is 
usually supplemented by very crude treatment of the final stage of 
the process, known as the freezeout. The spectra are
calculated by calculating Cooper-Frye integral over a certain surface, which in practice is
taken to be an isotherm with a particular temperature $T_\text{f}$. 

One of the improvements we try to develop in this work is to substitute
the isotherm by another surface, prescribed by a more meaningful freezeout condition
relating the expansion rate of matter to the corresponding reaction rates. 

As we will show below, the freezeout surface (FOS) consists of two distinct
parts. One of them, to be called the ``lid,"  is characterized by a Hubble-like flow, with transverse rapidity 
growing approximately linearly with the radial distance from the center, $\kappa\sim r$. This  part 
dominates spectra at not-too-large $p_\perp$, and was studied extensively, e.g.
with the so-called ``blast wave" parameterizations.  
While our work somewhat modified the FOS itself, the spectra at  $p_\perp< 2\, \text{GeV}$ remain unchanged as compared to multiple previous works. 
 
 The  part of the FOS we will discuss in this work is the ``outer edge." We will show that at larger  
 $p_\perp$, its contribution to the spectra becomes dominant.
 Indeed, in this region spectra are very sensitive to the  maximal value of transverse collective flow, 
 and the ``outer edge" generates transverse rapidity up to $\kappa\sim 1.4$ exceeding that of the ``lid" $\kappa<1.2$. 
 
 We will work out an analytic solution, approximating this region, based on the Riemann ``rarefaction fan" solution,
 and compare it with the standard numerical solution of hydrodynamics. We will further show that $T$ and $u^\mu$ in this region are indeed directly related to each other.
 
 We will also reformulate the freezeout condition itself, incorporating local information
 on the matter expansion rate  $\partial_\mu u^\mu$. Since even in the case of the rarefaction fan,
this quantity is {\em not} constant on the isotherm, we have to conclude that
 the correct FOS {\em cannot}  be 
 an isotherm. We will then show how the FOS gets modified.
 
 Accounting for all of this,
 we see rather encouraging description of particle spectra for a greater range of $p_\perp$.
 It is important to stress that this applies for secondaries of different masses, such as
 $\pi,K,N$. 
 
 In this paper, we will only discuss central collisions and axially symmetric flow. A
 paper to follow will deal with non-central collisions and elliptic flow.

\section{Motivation}
\subsection{The extent of thermal mass spectra at the chemical freezout}
\begin{figure}[htbp]
\begin{center}
\includegraphics[width=.5\textwidth]{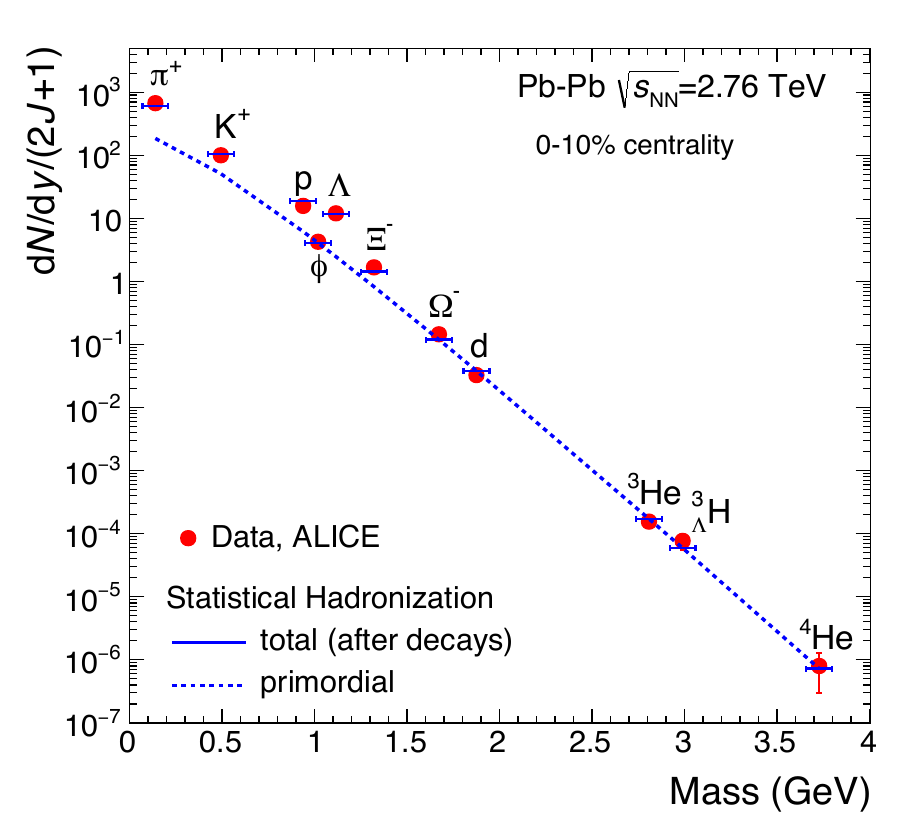}

\caption{ with the statistical weights  removed, plotted as a function of the particle mass.
The horizontal blue lines are with the decays of the resonances, the dotted blue line are
primordial only. From Ref. \cite{Andronic:2017pug}.
}
\label{fig_particle_ratios}
\end{center}
\end{figure}

A fireball created in heavy-ion collisions is generally considered to be a ``well-equilibrated"
thermal system, at least by its freezeout stages. A more precise meaning
of this statement comes from comparison of the mass distributions of the secondaries with
the Boltzmann exponent, corresponding to chemical freezeout temperature  
$T_\text{ch}\approx 156 \, \text{MeV}$ (and baryonic chemical potential $\mu_\text{b}$).  An example of such a
comparison  is replicated in Fig. \ref{fig_particle_ratios}, from Ref.  \cite{Andronic:2017pug}.
As one can see, the thermal description correctly reproduces the observed particle yields for 
 about nine orders of magnitude.

In this paper, we discuss not the mass but energy and momenta distributions. Those are determined at
the kinetic freezeout stage, corresponding to somewhat later times
as compared to chemical freezeout, with 
 lower temperature $T_\text{f}\approx 100\, \text{MeV}$ as compared to $T_\text{ch}\approx 156\, \text{MeV}$ . 
 It is widely expected that 
 collisions happening in a hadronic resonance gas, 
 between these two 
freezeout temperatures,  does not spoil the thermal equilibration.

Of course, 
particle masses are Lorentz scalars, and their distribution is unaffected by motion of matter, while the
particle energies and momenta are components of Lorentz vector and are affected by it. 
Unlike  the total particle yields, calculation of the observed spectra 
requires a precise knowledge of the flow velocity on the freezeout surface. 
But, with relatively small flow gradients at late stages of the collision, one may still
think that thermal distribution is  reasonably well maintained in each hydro cell.  
Optimistically, we therefore assume that with some distribution of the flow, either
obtained  from some 
phenomenological model or from numerical hydrodynamics, one may  be able to describe the particle 
  spectra over a similar range of $p_\perp$, or to energies (in co-moving frame) of a similar magnitude as the masses, or not less than $4\, \text{GeV}$. 

 \subsection{Looking for changes in the observed  $p_\perp$ spectra}
% origin of spectra and $v_n(p_\perp)$ below $p_\perp^{max}$ } 
%  understanding the main features   
  
Since we focus on large $m_\perp,p_\perp \gg T$ part of the spectrum in this work, 
the arguments of the Bessel functions in Eq. \ref{eqn_bessel2} are large, and 
one can approximate them by the exponent
with a ``blue shifted slope" $\hat T= T e^\kappa$,
\be \exp\left[-{p_\perp \over T}(\cosh(\kappa)-\sinh(\kappa))\right]=\exp\left[- {p_\perp \over \hat T}\right] \,.\ee

\begin{figure}[htbp]
\begin{center}
\includegraphics[width=.5\textwidth]{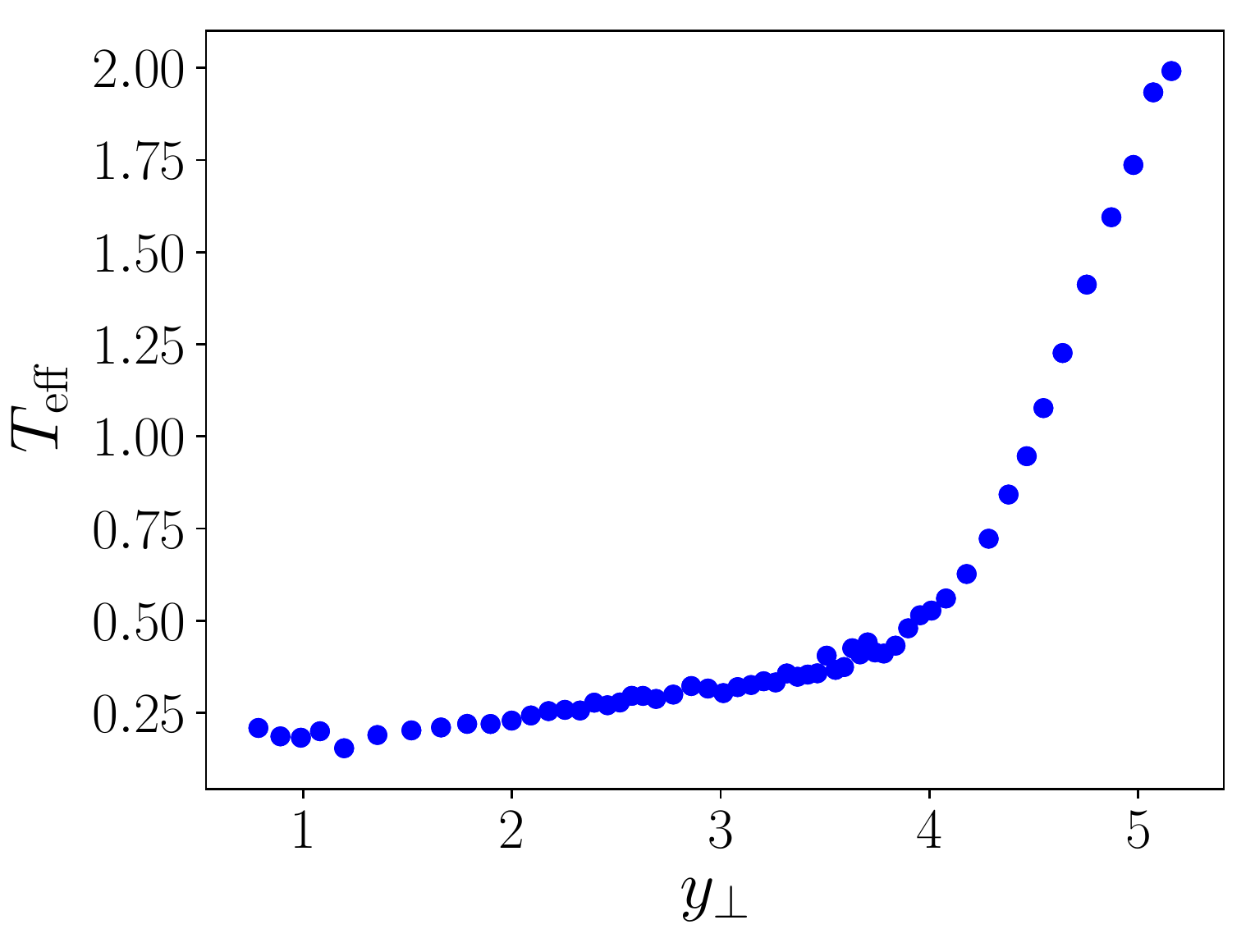}
\caption{The effective temperature defined as the logarithmic derivative of the $p_\perp$ 
distribution, as a function of transverse rapidity $y_\perp$.
}
\label{fig_pion_effT}
\end{center}
\end{figure}

Let us now have a look at the effective temperature extracted from the observed spectra, defined as its logarithmic derivative. This quantity is plotted in Fig. \ref{fig_pion_effT} as a function of transverse
rapidity $y_\perp=\text{arcsinh}[p_\perp/m]$. As one can observe from this plot,
the slope grows  very gradually,  till a value of  $ y_\perp \sim 4.2$ (corresponding to pion momenta $p_\perp>   5 \, \text{GeV/c}$). 
 At higher transverse rapidities, the slope changes and 
 starts to grow more rapidly.
 Similar changes of spectra are observed for other secondaries, at different rapidities but similar $p_\perp$.  
  This change of spectra, from near-exponential to power-like, is a well-known  phenomenon induced by hard partonic processes
related to jet production. In summary, spectra
themselves also suggest that perhaps the limits of thermal description
should be somewhere at $p_\perp \sim   5 \, \text{GeV/c}$, in the same ballpark as
the thermal distributions over masses. 

Now, how can pions reach transverse rapidity as large as $ y_\perp^\text{max}\sim 4$? 
What fraction of them are from thermal and what from collective motion? 
From
multiple studies of spectra at small $p_\perp$, we know that for central PbPb LHC collisions the value of kinetic freezeout temperature is rather low, $T_\text{f} \approx 0.1\,\text{GeV}$. Comparing it to the effective 
``blue shifted"
slope $\hat T\approx 0.4\,\text{GeV}$ of Fig. \ref{fig_pion_effT}, one may conclude that some parts of the FOS must have the ``blue-shift" factor $\exp(\kappa_\text{max})\sim 4$, corresponding to the
transverse 
  rapidity of the flow 
  %ranging between $\kappa\approx 0.8$ to its maximal value 
   \be \kappa_\text{max}\approx 1.5 \label{eqn_kappa_max} \ee
(The corresponding rapidity of the thermal motion in the matter frame should then be $y_\perp-\kappa_\text{max}\approx 2.5$, the
energy in the matter rest frame $E_\pi=m_\pi \cosh(2.5)\approx 0.85 \, \text{GeV}$, and the Boltzmann factor 
$\exp(-E_pi/T_f)\sim 2\times10^{-4}$, more than an order of magnitude above the fraction of the transverse momentum spectrum at the discussed $p_\perp$ of $\sim 10^{-5}$.)

 One of the 
main points of this paper, is that the ``outer edge" of the fireball can in fact deliver a transverse collective flow of this magnitude.   
  
 The questions we will be discussing below are then: 
 \begin{enumerate}[(i)]
\item Can the collective flow with the transverse rapidity as large as $\kappa\sim \kappa_\text{max}$
be phenomenologically acceptable, providing consistent description of  spectra for species of secondaries of very different mass? 
\item Can  a  flow with a transverse rapidity as large as $\kappa\sim \kappa_\text{max}$ be  generated hydrodynamically?
\item Where would this region of the fireball be located? Can it be analytically understood?
% in some approximation? 
\end{enumerate}
  We will argue that all of them can be answered in the affirmative.
  
\section{Using analytic solutions of relativistic  hydrodynamics  }  
  \subsection{Gubser flow}
 While mainstream
development of these ideas relied mostly on numerical codes, without and with viscosity terms,
one cannot overestimate the value of analytic solutions, ranging from the
original Landau solution \cite{Landau:1953gs}, the Bjorken rapidity independent 1+1 dimensional solution \cite{Bjorken:1982qr} and the Gubser solution \cite{Gubser:2010ze,Gubser:2010ui}, which makes use of
scale invariance of quark-gluon plasma (QGP) and is obtained by a conformal transformation.

Apart of pedagogical value, and that of code verification, analytic solutions
allow us to reach understanding and derive results, which would otherwise 
take a much longer time. For example, using small perturbation on top of Gubser flow,
P. Staig and one of us \cite{Staig:2011wj} found a complete Green function -- expanding 
deformed sound sphere from a delta-function-like source in the initial conditions --,
which predicted a very non-trivial shape of two-particle correlation function in azimuthal angle,
soon confirmed by experiment. 
Gubser flow has also been used for calculation of the radial flow \cite{Shuryak:2013ke} and femtoscopy
radii \cite{Hirono:2014dda} 
in ``small systems,"
the high-multiplicity $pp,pA$ collisions.

These studies, however, have revealed that the Gubser solution can resemble real
 heavy-ion collisions, and thus be useful, only for the ``lid"
   portion of the freezeout surface. This is because of its
 relatively slow (power-like) decrease of the densities with distance, which
  do not correspond to actual fireballs. One may explain it as follows: Gubser's fireball expands not in vacuum, but 
  into some kind of atmosphere, which prevents the necessary flow development at the outer edge.

%In this paper we show that, between the dense matter in the fireball and the vacuum outside it,
%there should be a universal region known as ``rarefaction fan", and then study 
%study whether secondaries coming from it can (or cannot) do indeed populate the
% intermediate region of transverse momenta mentioned above.

  Looking for solutions independent on both ``angles" $\eta,\phi$ out of four coordinates,
  Gubser's solution is 
\begin{eqnarray}
u_{\mu} &  = &
\left(-\cosh{\kappa(\tau,r)},0,\sinh{\kappa(\tau,r)},0\right)
\end{eqnarray}
Gubser obtained the following solution
\begin{eqnarray} \label{G_v}
v_\perp & = & \tanh{\kappa(\tau,r)}  =  \left(\frac{2q^2\tau
r}{1+q^2\tau^2 + q^2r^2}\right)
\end{eqnarray}
\begin{eqnarray}   \label{G_energy}
\epsilon & = & \frac{\hat{\epsilon}_0 (2
q)^{8/3}}{\tau^{4/3}\left(1+2q^2(\tau^2 +
r^2)+q^4(\tau^2-r^2)^2\right)^{4/3}} \nonumber
\end{eqnarray}
The solution has two parameters, $\hat{\epsilon}_0$ and $q$, representing the
scale of the energy density and the size of the fireball, respectively. 
  
\begin{figure}[htbp]
\begin{center}
\includegraphics[width=.5\textwidth]{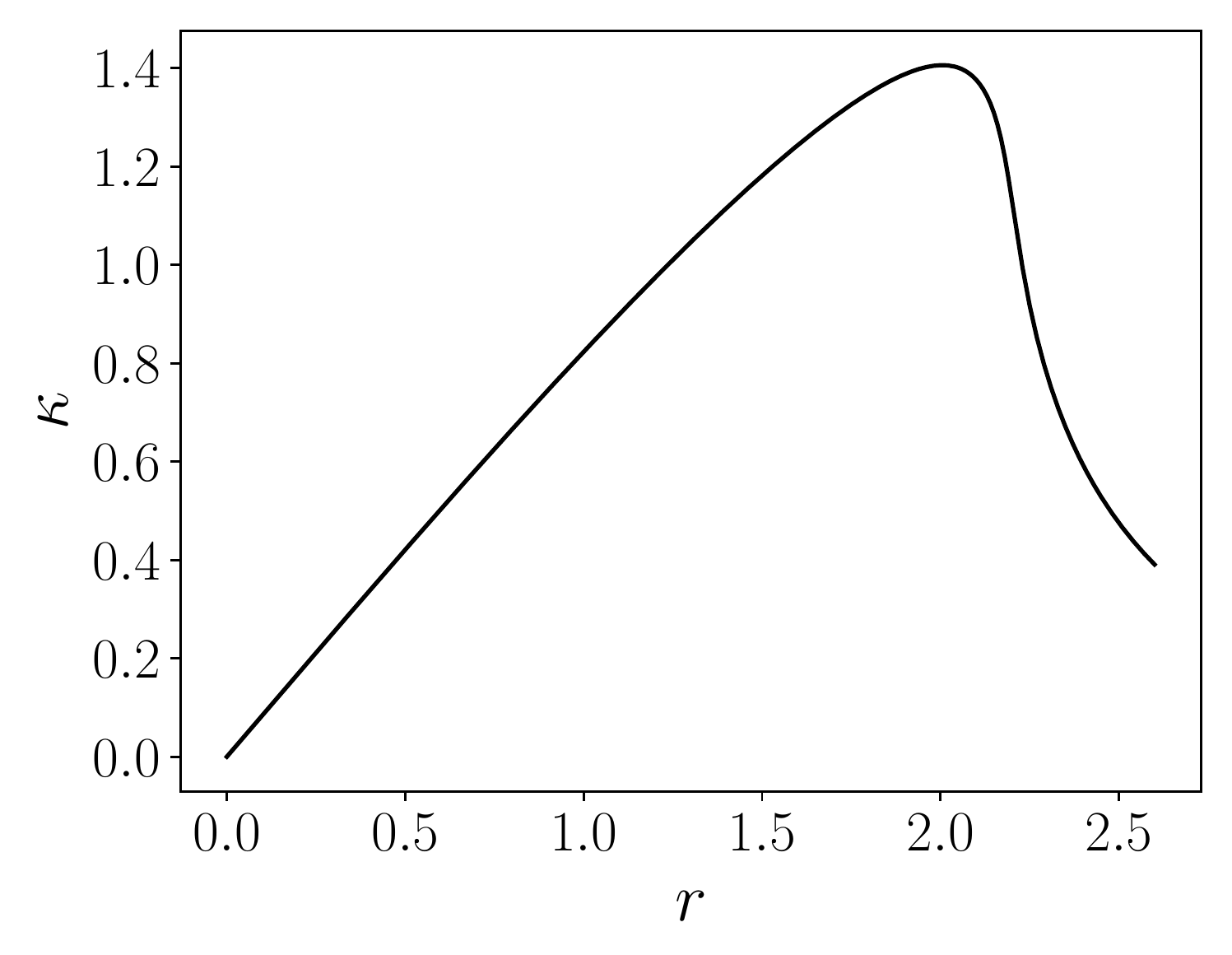}
\caption{The value of the transverse flow rapidity $\kappa$ versus
the radial distance $r$ (in $1/q$ units) at the freezeout isotherm for Gubser flow.}
\label{fig_kappa_Gubser}
\end{center}
\end{figure}

  Requiring the freezeout surface to be the isotherm $\epsilon(\tau,r)=\epsilon_\text{f}$
  one can find $\tau_\text{f}(r)$, and, substituting it 
  into the collective velocity/rapidity, find its distribution on the surface. An example of such 
  procedure is shown in Fig.\ref{fig_kappa_Gubser}. 
  
  The linear growth from $r=0$ corresponds to the ``lid" part of the freezeout surface.
  At the peak of this distribution
  the value of the   transverse flow rapidity $\kappa$ may be tuned to a phenomenological value
  needed to describe the data, $ \kappa_\text{max} $, yet it does so only
 at the thin ``rim" of the fireball. At larger $r$,  one finds that 
  the flow rapidity {\em decreases} with distance.   We will argue that this behavior is wrong, as it is inconsistent with the trend  expected from universal properties of ``rarefaction fan" portion
  of the solution and actual numerical solutions. This fact has been emphasized already in Ref.
  \cite{Staig:2011wj}, and traced to the unphysically slow (power-like) decrease of the initial matter
  distribution. (To give the idea of what such slow tail means, compare it to an explosion
  in the atmosphere, which stops the flow at certain distance.) Therefore, spectra were calculated in Ref. \cite{Staig:2011wj} using only the late-time part of the freezeout surface.  
  
   For the record, one can calculate the expansion rate for Gubser flow, and find it to be
   (in $q=1$ units) 
   \be \partial_\mu u^\mu= {4 \tau \over \sqrt{(1+r^2+\tau^2)^2-4r^2 \tau^2 }} \ee
Using it, one can plot the surfaces corresponding to the freezeout condition.%  and
%  flow rapidity  
%  mist  increase indefinitely  $\kappa\rightarrow \infty$ at the light cone, bordering with vacuum.
%It is the derivation of the correct surface flow which will allow us to describe the
%observed spectra at higher $p_\perp$. 
  
%%%%%%% needed ?  
%In this section will be focusing on the region of relatively small time and distances $r,\tau \ll 1/q$, in which both denominators can be approximated by unit value. The transverse velocity/rapidity is simply following ``the double Hubble" expansion $v_\perp\sim \tau \, r$, so the lines of constant flow velocity on the $\tau-r$ plane are hyperbolae $\tau \sim 1/r$. The energy density in this approximation is $\epsilon \sim 1/\tau^{4/3}$ independent of $r$: thus the isotherms are horizontal lines  on the $\tau-r$ plane. It is important that $v_\perp=const$ and $\epsilon=const$ lines are very different. 
%The very important role in forming the spectra is thus played by the region close to the 
%``corner" of the freezeout surface ($T=T_f$) at which  $v_\perp$ is maximal. 

%%%%%%%%%%%%%%%%%%%%%%%%%%%%
   
\section{The rarefaction fan} 

%The specific solution we will give below in this section, and its explicit derivation 
%in Appendix. Let us however start with explaining what is so qualitatively different
%in this region, relative to the flow on the later side of the freezeout surface. 

The solution we discuss is, in principle, known; in the fully-relativistic 1+1d context,
it was first discussed  in Ref. \cite{Baym:1984sr}.
However, at that time, the magnitude of the flow was expected to be 
very small: e.g. in that paper, pions with $p_\perp\sim 30\,\text{MeV}$ were considered,
which is about hundred times (!) smaller than what we are going to discuss below.  
 We will also rederive the solution, in pedagogically simpler way.

% The rarefaction fan solution has been studied in our field by Rischke et al
% The initial conditions used, $\epsilon=const, r<R$ and $\epsilon=0,r>R$
% led to the so called  ``burning log" scenario, in which no explosion 
% happened because in the most part of fireball gradients of pressure was assumed to be zero. 
% Nothing like that happens if one uses the realistic initial conditions: in fact
% the hydrodynamical explosion is very robust, as hydrodynamics predicted and RHIC experiments confirmed.  

The main simplification, leading to the expansion solution we will discuss, is due to Riemann,
who assumed that
 in the rarefaction fan region, the two unknown functions, e.g.
 the energy density $\epsilon(\tau,x)$ and flow rapidity $y(\tau,x)$,
are {\em not} independent but in fact  are directly related to each other, namely
\be  \epsilon(\tau,x)=f\big( y(\tau,x) \big) \,.\ee 
Therefore, the isotherms, which are not exactly the FOS but close to it, and $y(\tau,x)=\text{const.}$
are the {\em same lines} in this region!
This property has the historic name of ``Riemann simple wave."
(In fact, many numerical solvers for astrophysical relativistic hydro are based 
on application of it at each space cell and each time step.)

 Now is perhaps the time to revisit its usage, applying it now to where it belongs,  the large-momentum
 tail of the particle spectra. 
Here, for pedagogical reasons,  we give explicit solution of the rarefaction fan 
directly from the equations, without use of a Riemann invariant.

For 1+1 dimensional relativistic hydrodynamics, we will not change coordinates, keeping the 
original $t,x$ coordinates. Using standard rapidity notations for 4-velocity,
\be u^0=\cosh(y), \,\,u^1=\sinh(y)\,, \ee 
one has the stress tensor in the form
\ba T^{00}=\epsilon\, \cosh^2(y)+p \,\sinh^2(y) \,, \label{eqn_EOM}
\\
T^{01}=(\epsilon+p) \cosh(y)\sinh(y)\,, \\
 T^{11}=\epsilon \,\sinh^2(y)+p \,\cosh^2(y)\,, \ea
subject to two equations
\ba {\partial T^{00} \over \partial t}+  {\partial T^{01} \over \partial x}=0 \,,\\
 {\partial T^{01} \over \partial t}+  {\partial T^{11} \over \partial x}=0 \,.\ea
For simplicity, we will use EOS $p=c_s^2 \epsilon$ with
constant speed of sound $c_s$. 

The first step toward solving the above is the
 Riemann idea to look for a solution in which all unknown functions 
 are directly related to each other, in our case
 \be \epsilon(t,x)=F\big( y(t,x) \big) \,.
 \ee
 The second idea is that, since the EOS used has no dimensional parameters, the
 solution is perhaps self-similar
 \be y(t,x)=f\left({ t \over x }\right) \,.
 \ee
 \begin{widetext}
 Substituting those to the equations of motion (Eq. \ref{eqn_EOM}) one have them in the form
 $$
{F'(f) \over F(f)}+  { -2 (1 + c_s^2) (t \,\cosh(2f) - x \,\sinh(2f) )\over 
  x - c_s^2 x + (1 + c_s^2) x \, \cosh(2f) - (1 + c_s^2) t \, \sinh(
     2f))}  =0 \,,$$
$$   {F'(f) \over F(f)}+  {2 (1 + c_s^2) (x \, \cosh(2f) - t \,\sinh(2f)) \over 
 t - c_s^2 t - (1 + c_s^2) t \,\cosh(2f) + (1 + c_s^2) x \, \sinh(
    2f) }  =0 \,.
 $$
 \end{widetext}
 The key consequence of the Riemann assumption is that the derivative  $f'$ appears in all terms and, if nonzero, can be cancelled out. As a result, one can find $f$ from the ordinary (not a differential)
 equation, requiring two complicated ratios in the two EOM to be equal to each other.
 This relatively complicated equation leads to a surprisingly simple answer,
 \be  y=f(t/x)= {1 \over 2} \log\left[\left({1-c_s \over 1+c_s}\right)\left({t+x \over t-x }\right) \right] \,. \ee
 Note that the logarithm in the second bracket corresponds to the so-called spatial rapidity
 \be \eta(t,x)={1 \over 2} \log\left({t+x \over t-x }\right) \,. \ee
Substituting it into these ratios, one finds that both are indeed the same and the remaining equation takes the form
$$ {1+c_s^2 \over c_s}+  {F'(f) \over F(f)}=0\,,$$
from which $$F(f)=\exp\left[-\left({1+c_s^2 \over c_s}\right)f \right]\,.$$
times an arbitrary constant, to be determined from the boundary conditions.

For discussion to follow, we will need the scalar expansion rate, so let us give it for
this flow
\be \partial_\mu u^\mu={1 \over  \sqrt{(1-c_s^2) (t^2-x^2)}} \,, \ee
which is, as expected, Lorentz invariant. 

%%%%%%%%%%%%%%%%%%%%%%%%%%%%%%%%
\section{The ``conical cup" model}
After discussion of the Gubser solution, we would like to introduce a simple model
which has properties much closer to what realistic hydrodynamical explosion predicts.

\begin{figure}[htbp]
\begin{center}
\includegraphics[width=7cm]{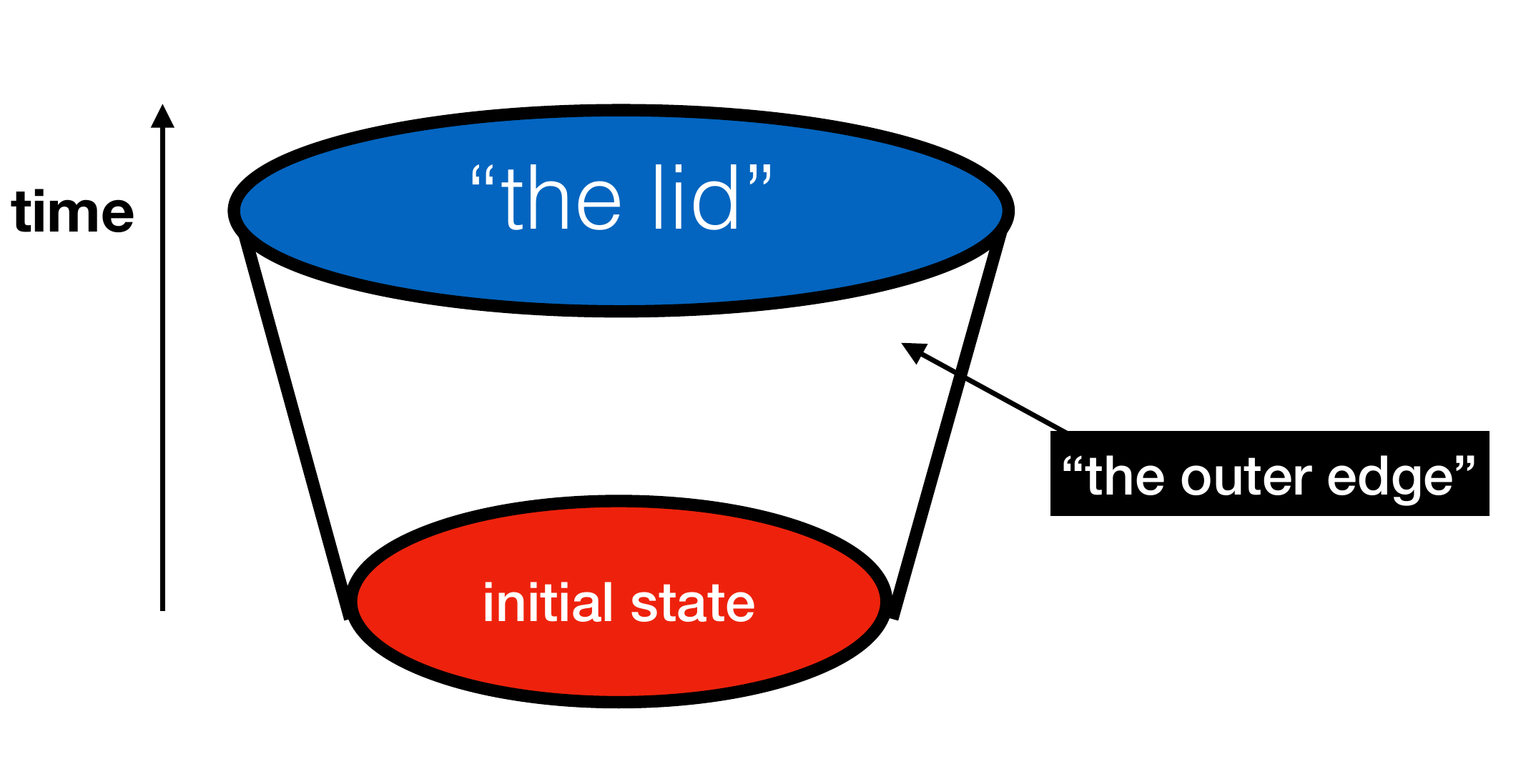}
\caption{Schematic picture of the freezeout surface in coordinates $r$-proper time $\tau$. }
\label{fig_cc}
\end{center}
\end{figure}

The schematic shape, we call the ``conical cup", is shown in Fig.\ref{fig_cc}.
It consists of the ``top" part, taken to be at constant (longitudinal) proper time $\tau=\tau_\text{f}$,
and the ``outer edge", taken to be of conical shape, with radius linearly
interpolating between the initial and final radii, $R_\text{i}$ and $R_\text{f}$. The bottom part is where the
initial conditions are to be defined; it is understood to be at small but finite proper time $\tau_\text{i}$
where one can start hydrodynamical description. We will not discuss the bottom part in this 
paper anymore, and the time $\tau_\text{i}$, typically a fraction of fm/c, is simply neglected
compared to $\tau_\text{f}\sim 17\, \text{fm/c}$. For justification of the model and realistic values of its parameters,
the reader should wait for later sections.

 We further assume that at the top part of the surface the transverse rapidity 
 dependence on $r$ is linear, up to its maximal value
 \be \kappa(r)= \kappa_\text{max} {r\over R_\text{f}} ,\,\,\, (r<R_\text{f}) \ee    
 Note that because $\tau=\text{const.}$ on the top part, the second term in the preceding expression vanishes.
 
 On the outer edge, we assume that the  transverse rapidity takes the same value on the whole wall, 
 $   \kappa =\kappa_\text{wall} $. (Justification to follow later.) If so, the  Bessel functions
 decouple from the integral over the wall, up to the volume factor
 \be V_\text{wall}=\int_0^{\tau_\text{f}} d\tau  \pi R_\tau^2\,. \ee
 As we will show later, only a part of the wall actually corresponds to the 
 solution with the constant (the highest)  $\kappa=\kappa_\text{max}$, so we introduce
 additional factor $P_\text{wall}$ to the outer edge contribution.
  Note also, that in this case the two terms with different Bessel functions  
 compete.
 
 \begin{widetext} 
 The resulting spectra for this model are shown in Fig. \ref{fig_cc}, for the pions and protons.
 The previous studies focused on the ``lid" contribution, shown by dash-dotted lines, which provides good description  of the shape of the distribution up to $p_\perp\sim 3\, \text{GeV}$.
 Note further, that the excess seen in the data at small $p_\perp$ is due to the so-called ``feed-down"
 contribution, that of the decay of multiple mesonic and baryonic resonances. It is well known
 and calculated in statistical hadronization models, such as Ref. \cite{Andronic:2017pug}.  
 
The new contribution of the ``outer edge" we introduced now, shown by the solid lines,
is small in the integral, of the order of a percent. However it becomes dominant at higher
transverse momenta, improving agreement for pions for $p_\perp=2-5\, \text{GeV}$ and for protons 
for  $p_\perp=3-6\, \text{GeV}$.  It significantly extends the description of the spectra in terms of absolute probability, by 2-3 orders of magnitude. This is important, because it 
now extends right  up to the momenta
$p_\perp>5-6\,\text{GeV}$ where the contribution from hard partonic reactions, eliminating the ``nobody's land" region in between. 

 \begin{figure}[h]
\begin{center}
\includegraphics[width=5.5cm]{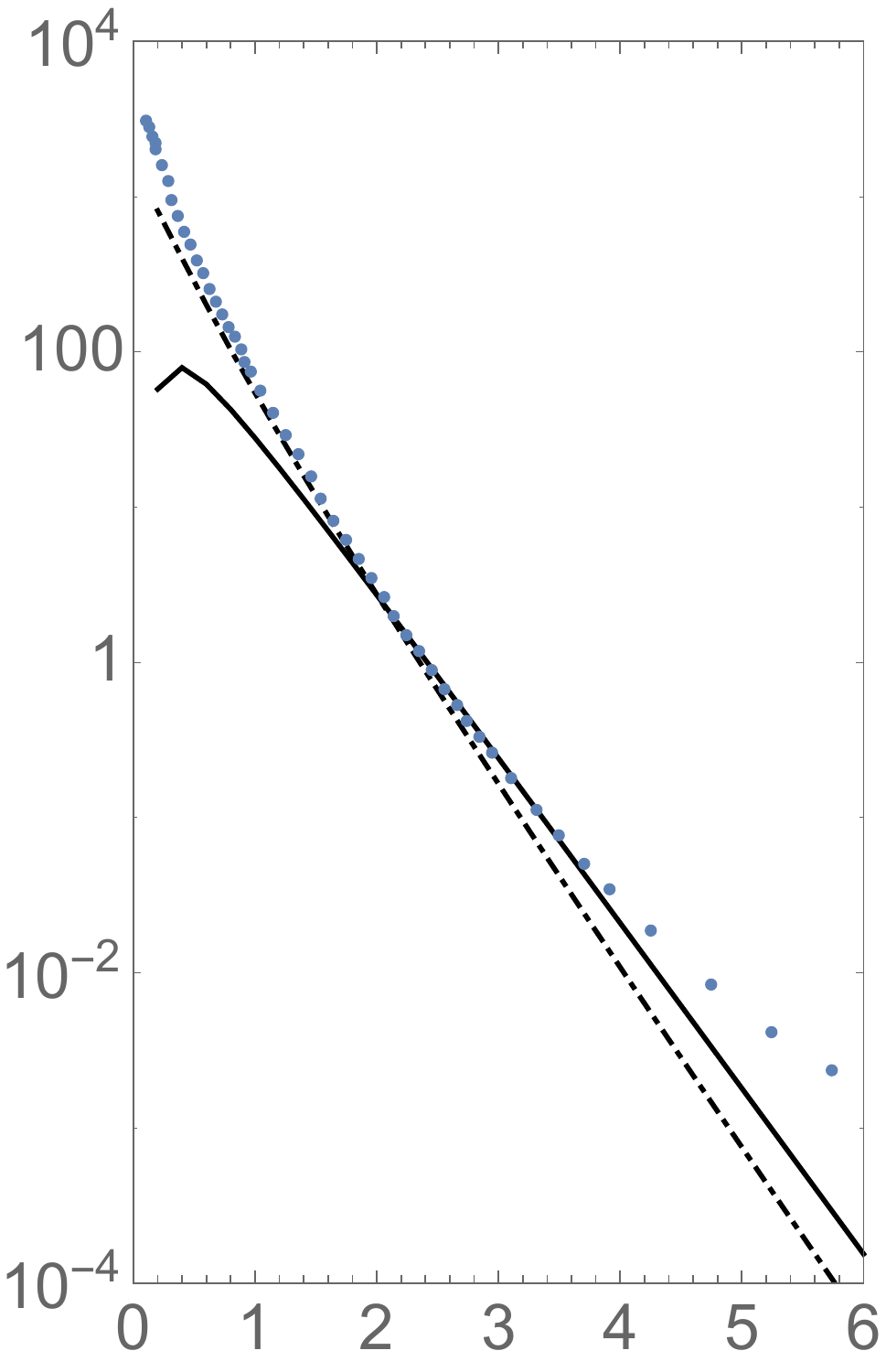}
\includegraphics[width=5.5cm]{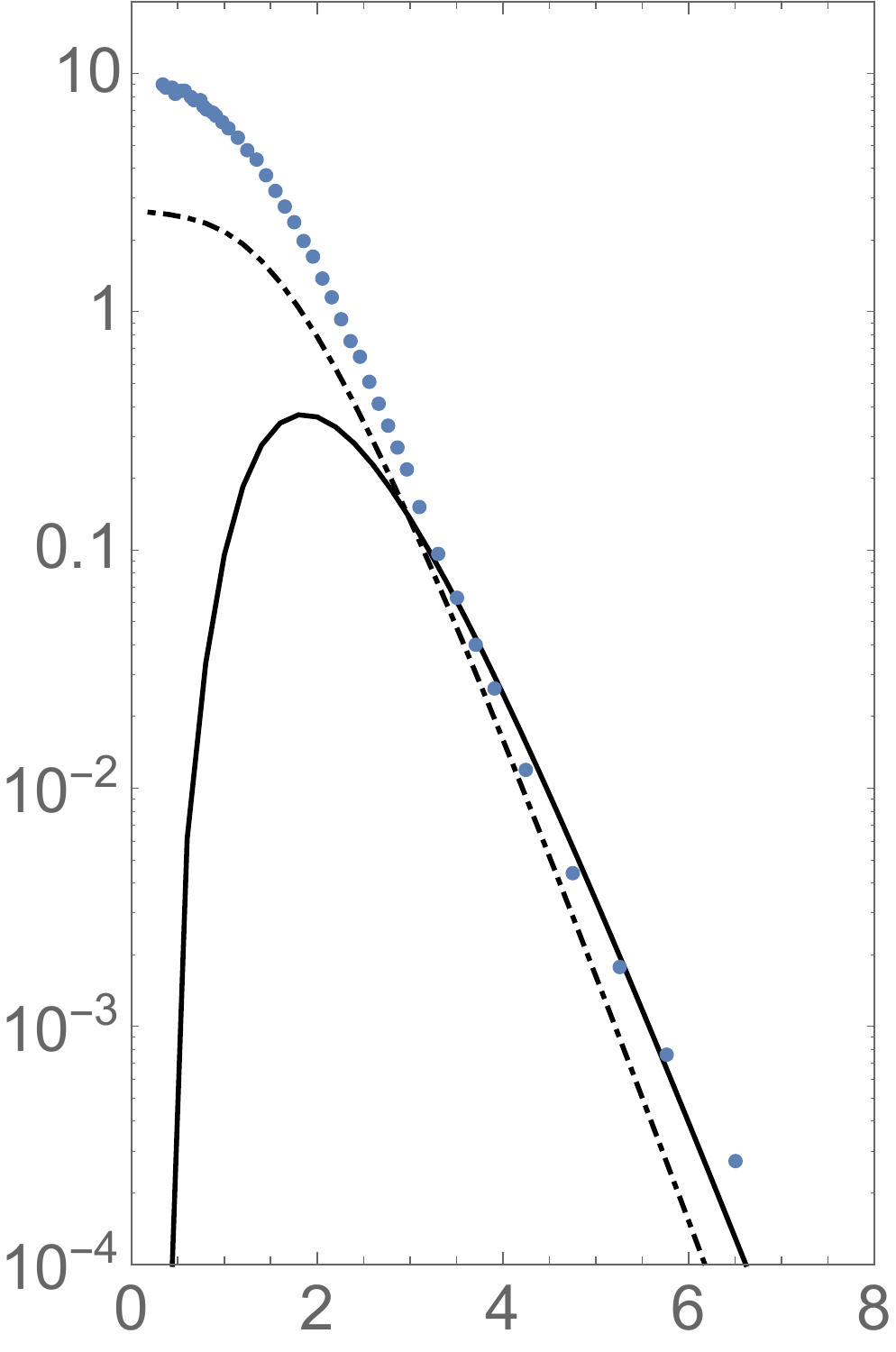}
\caption{The transverse momentum spectrum $dN/dy dp_\perp^2 \big(\text{(GeV/c)}^{-2}\big)$
versus transverse momentum $p_\perp\, (GeV/c)$
for pions (the left plot) and protons (the right plot). The points correspond to ALICE 0-5\% 
centrality data from Ref. \cite{Abelev:2014laa}, the dash-dotted lines show the ``lid" contribution,  and 
the solid lines show that of the ``outer edge," with additional factor  
$P_\text{edge}=1/4$.}
\label{fig_cc}
\end{center}
\end{figure}
 \end{widetext}
 
%%%%%%%%%%%%%%%%%%%%%%%%%%%%%%%%%

\section{Transverse flow at the isotherm surfaces of numerical hydrodynamics} 

As input, we use an ensemble of hydrodynamic solutions generated from
Glauber-based initial conditions by the MUSIC \cite{Ryu:2012at}.  We use smooth (Glauber-based) initial conditions and
take small impact parameter, corresponding to 0-5\% centrality bin, the same as
in the ALICE data \cite{Abelev:2014laa} to which we will be comparing our results. 

\begin{figure}[t]
\begin{center}
\subfigure[]{
\includegraphics[width=.48\textwidth]{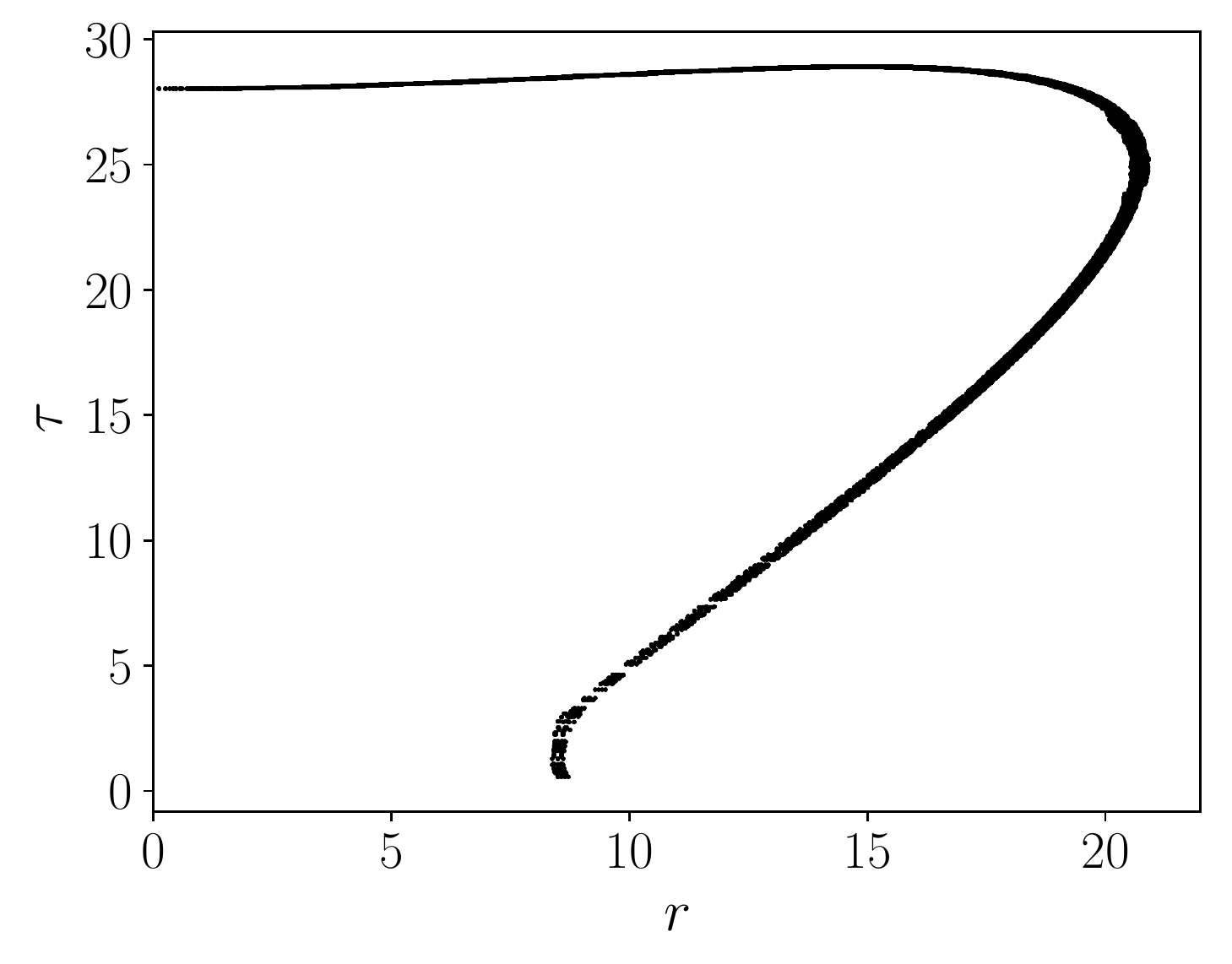}}
\subfigure[]{
\includegraphics[width=.48\textwidth]{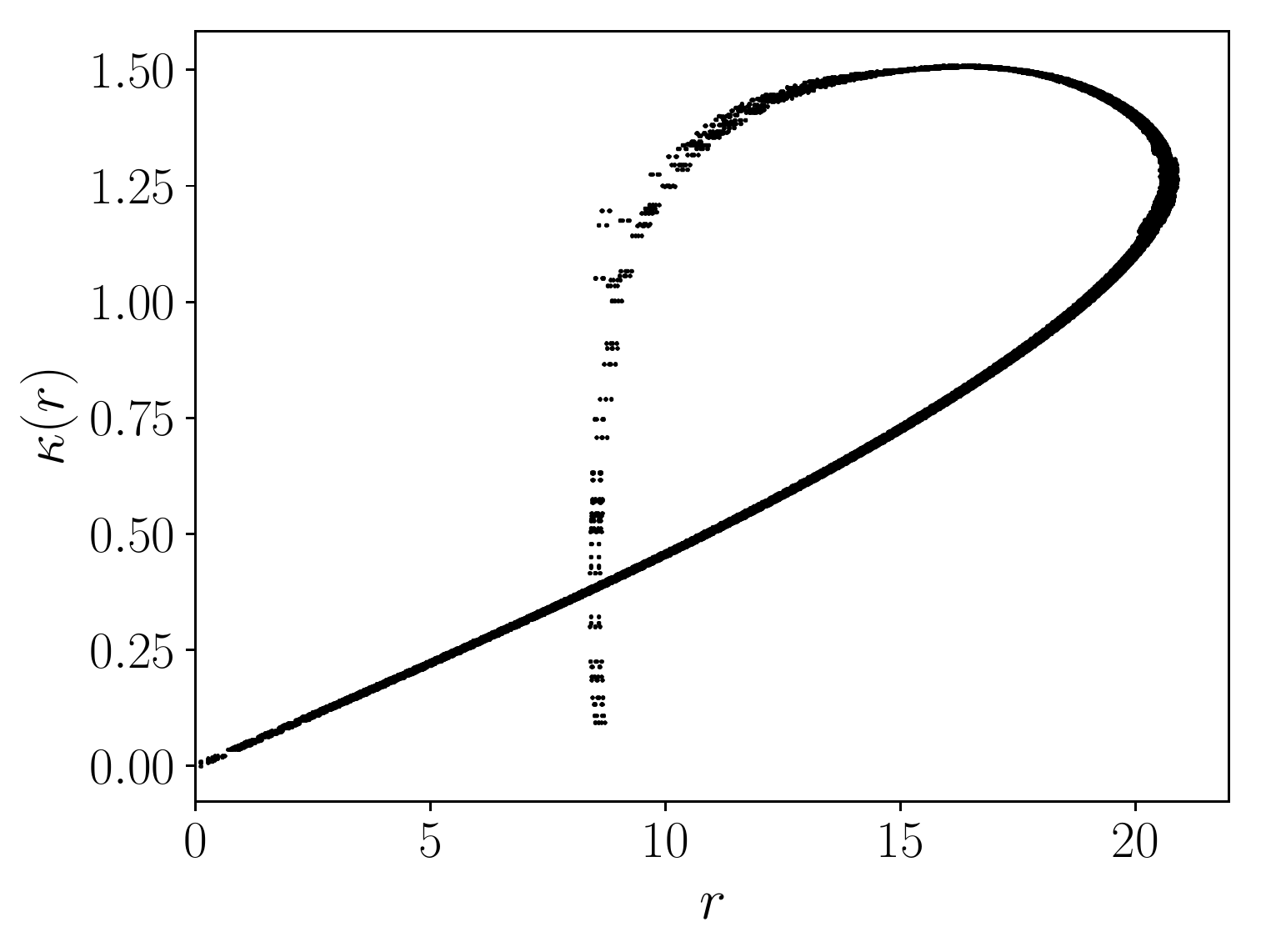}}
\caption{The upper plot (a) shows the freezeout surface $T=100\, \text{MeV}$, in $r,\tau (\text{fm})$ coordinates. 
The lower plot (b) shows the transverse flow rapidity $\kappa(r)$ versus the radial distance $r (\text{fm})$. The dots represent hydro cells at the freezeout surface, corresponding to numerical solution using MUSIC code for 0-5\% centrality.}
\label{fig_rapidity_at_fo}
\end{center}
\end{figure}

In the upper Fig. \ref{fig_rapidity_at_fo} we show location of the points at the isotherm $T=T_\text{f}=100\, MeV$, on the $r-\tau$ plane. Their distribution clearly display two 
parts of the f.o. surface: (i) the ``lid;" and (ii) 
the ``outer edge," discussed above. 
%Note that the former is close to having constant proper time $\tau_{f.o.}\approx 27\, fm/c$. Note further than the ``outer walls" have very small vertical (that is, no flow) part at $\tau<3\,fm/c$, and then
% extends linearly, from the initial edge of the nuclei at $r\approx 7\, fm$ to about 
% $r\approx 20\, fm$. 
The slope of the edge tells us that it moves outward with the velocity
 $v_\text{edge}\approx 1/3$ (not to be confused with the collective flow velocity on this edge,
 which is much larger). 

In  the lower plot  of Fig. \ref{fig_rapidity_at_fo}, we show the corresponding distribution of 
the transverse flow rapidity
 $\kappa(r)$. It consists of a familiar Hubble-like linear rise on the ``lid"
part of the FOS, complemented by  an upper ``loop," generated at the ``outer edge". This plot is to be contrasted with Fig. \ref{fig_kappa_Gubser} for the Gubser flow, in which such  a ``loop"
is absent. The main observation here is that the ``loop" indeed reaches the maximal
transverse rapidity value (Eq. \ref{eqn_kappa_max}) discussed above, required to describe the high-$p_\perp$ spectra.

% we show  the distribution of the rapidity of the flow $y_\perp$. The light histogram correspond to the ``lid" ($\tau>17\, fm/c$), and the dark one to the ``outer edge". While the former shows a remarkably flat distribution, in the range $y_\perp\in [0.1,1.1]$,
%the latter has a clear peak at  $y_\perp\approx 1.2$. The existence of this peak
%is the consequence of near-Riemann nature of the hydro solution at the edge,
%according to which the values of the temperature and rapidity must be strongly correlated. 

%\begin{figure}[t]
%\begin{center}
%\includegraphics[width=7cm]{freezeout_r_kappa}
%%\includegraphics[width=7cm]{rapidity_at_fo}
%\caption{The transverse flow rapidity $\kappa$ versus $r (fm)$, dots are cells at the freezeout surface. %(lower plot) The histogram of the  transverse flow rapidity $\kappa$  distribution over the freezeout surface. The lighter part corresponds to the ``lid" region,
%%defined as $\tau>27, \,fm/c$, the darker part shows the contribution of the ``outer edge", at  $\tau < 27, \,fm/c$}
%}
%\label{fig_freezeout_r_tau}
%\end{center}
%\end{figure}

  \section{The improved  freezeout condition and FOS's} 
 For high energy collisions, the idea of freezeout \cite{Pomeranchuk:1951ey}
  predated even the first application of relativistic hydrodynamics by Landau.
  It is also widely used in Big Bang cosmology. 
  Schematically, the condition reads 
  $$ \text{collision rate} = \text{expansion rate} $$ 
  A more precise form used in Ref. \cite{Hung:1997du} is, for the particle of type $i$
  \be  \dot{w}_i=\langle \sum_j n_j \sigma_{ij} v_{ij} \rangle \approx \partial_\mu u^\mu \,,\ee
  where in the l.h.s. the sum is taken over all particle types $j$, with mutual scattering cross section
  and relative velocity. Note that the collision rates in the l.h.s. depend on $i$;
  the freezeout surface is not the same for different particle types. For example, the nucleons
  have large $\sigma_{\pi N}\approx 200 \, \text{mb}$ at the $\Delta$ peak, and thus they
  are expected to decouple {\em later} than the pions. Strange particles such as $\phi$ 
  have smaller cross sections, and thus should decouple {\em earlier} than the pions, etc. 
  
  Although these considerations  and the collision rates in hadronic gas are well known, we have not seen them being used often, since that early paper \cite{Hung:1997du}. 
  
  The rates  generally  grow rapidly with the temperature $T$, and can be conveniently parameterized in a power form    \be \dot{w}_i=C_i T^{P_i}\,. \ee
  For example, in a cool gas of pions described by the Weinberg chiral Lagrangian, one finds
  $ \dot{w}_\pi \sim T^5/f_\pi^4$.   Taking the root of
  the power $P_i$ and rewriting the freezeout condition back for $T$, one can put it in the following convenient form
  \be T(x)= \left[{ \partial_\mu u^\mu(x) \over C_i}\right]^{1/P_i} \,.  \label{eqn_FO_condition}\ee
So, when the power $P_i$ is very large, the  r.h.s. is nearly constant, and
thus the f.o.surface reduces back to  the isotherm. However, the realistic power
is $P\approx 3.5$, and  in general both $T(x^\mu)$ and $\partial_\mu u^\mu(x^\nu)$ are some nontrivial functions
of the space-time coordinates, obtainable from hydro equations.

Before presenting the results of the numerical hydrodynamics, 
let us now analyze the  space-time distribution of expansion rate, starting from
the ``lid", and then proceeding to the ``outer edge".

Taking the ``lid" at fixed proper time and taking, for simplicity, the non-relativistic part of it,
one can express linear rapidity growth with $r$ as a ``Hubble flow"
\be v_r=H r \ee
The divergence of this, in spherical coordinates, is
\be \partial_m u^m\approx {1 \over r^2} {\partial \over \partial r} (r^2 v_r) =3H \,, \ee
which is a constant. Its magnitude is readily obtained; from the slope of the rapidity plot,
one finds that $H\approx 1/(20\, \text{fm}) $ from which $(1/\tau_\text{exp})=3H\approx (1/7\, \text{fm})$.
This value agrees well with the pion collision rate $1/\tau_\text{coll}(T=100\, \text{MeV})$ from
Ref. \cite{Hung:1997du}.

On the ``outer edge," a similar estimate goes as follows. In this case, one should use
cylindrical coordinates, so the expansion rate is
\be {\partial \over \partial \tau}\cosh(\kappa)  + {1 \over r} {\partial \over \partial r} (r \sinh(\kappa)) 
\ee
It is simple to calculate it at the isotherm, since according to rarefaction solution,
on the isotherm, the flow rapidity is constant. Therefore, the first term is zero, and the second gives simply
\be {1 \over \tau_\text{exp}}\approx {\sinh(\kappa) \over r} \,,\ee      
and is therefore {\em not} constant over the outer edge region.
This agrees with conclusion from analytic 1+1d solution above, for which $\partial_\mu u^\mu\sim 1/\tau$, and means that the FOS in the outer edge region needs to be modified. 
 
%Let us see by how much one should deform it. 
%According to Fig.\ref{fig_freezeout_r_tau}(a), the region
%in which $\kappa\approx const$
%corresponds to distances from $r=9\, fm$ to about $r=18\, fm$,
%  the expansion rate changes on it by the factor two. This factor, taken 
%  in the power $1/P$ in the freezeout condition (\ref{eqn_FO_condition})  
%   is still relatively constant: e.g. for Weinberg pion gas with $P=5$ it only changes by 
%   $15\%$. So, at smaller $r$ one needs to shift to $T_f$ $15\%$ larger, or $T\approx 115\, MeV$ instead. The  rapidity of the collective expansion $\kappa$ would therefore be
%   smaller, by about ???. 
%   
%   Summarising: In the ``lid" area the isotherm and lines of constant expansion
%   rates simply coincide. In the  ``outer edge" region the r.h.s. of the freezeout condition (\ref{eqn_FO_condition}) is not constant, but varies in reasonably small interval. 
%   Corrections to the FOS can be readily obtained perturbatively. 
%
%Plot of expansion rate 

\begin{figure}
\begin{center}
\subfigure[]{
\includegraphics[width=.48\textwidth]{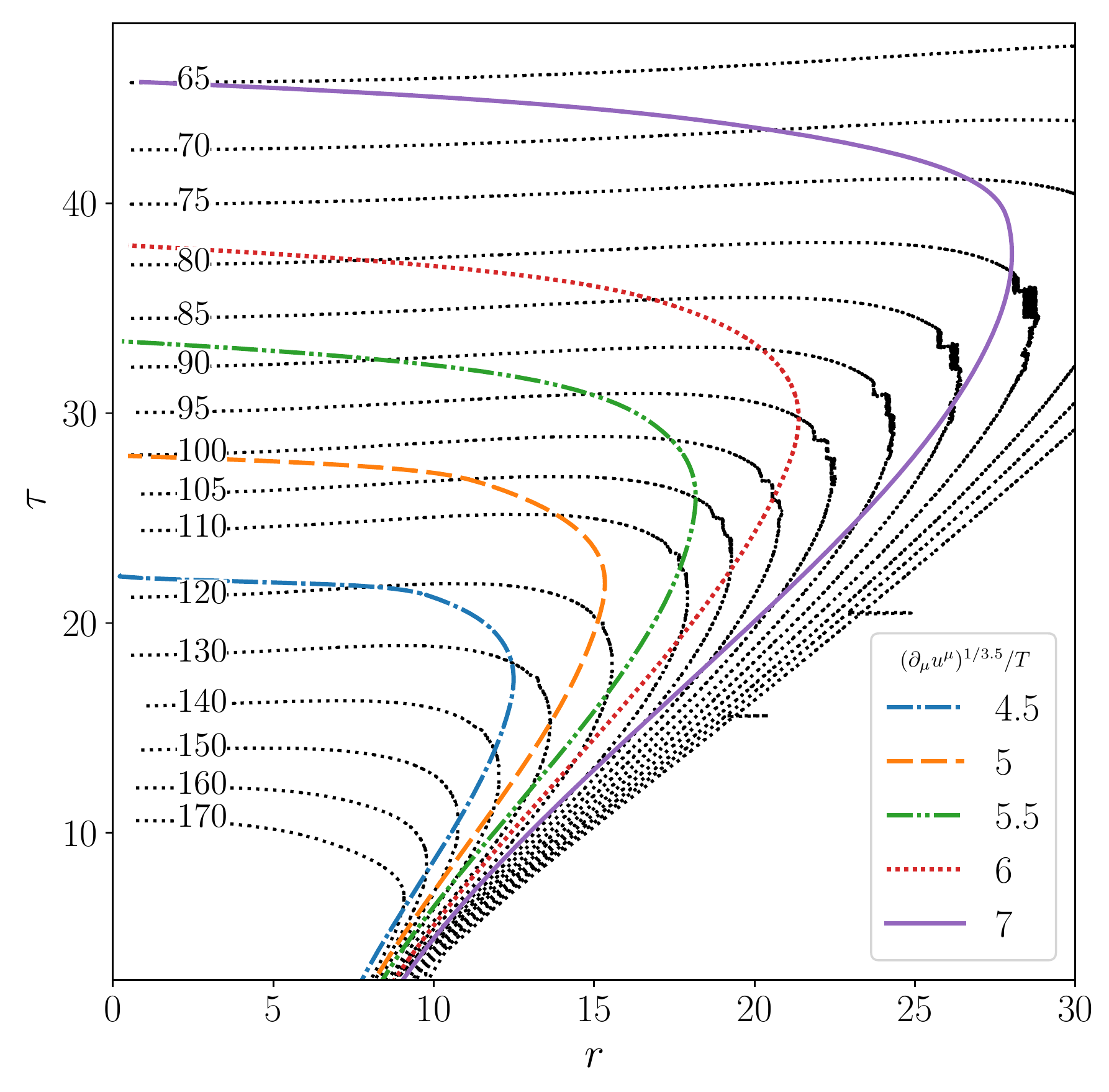}}
\subfigure[]{
\includegraphics[width=.48\textwidth]{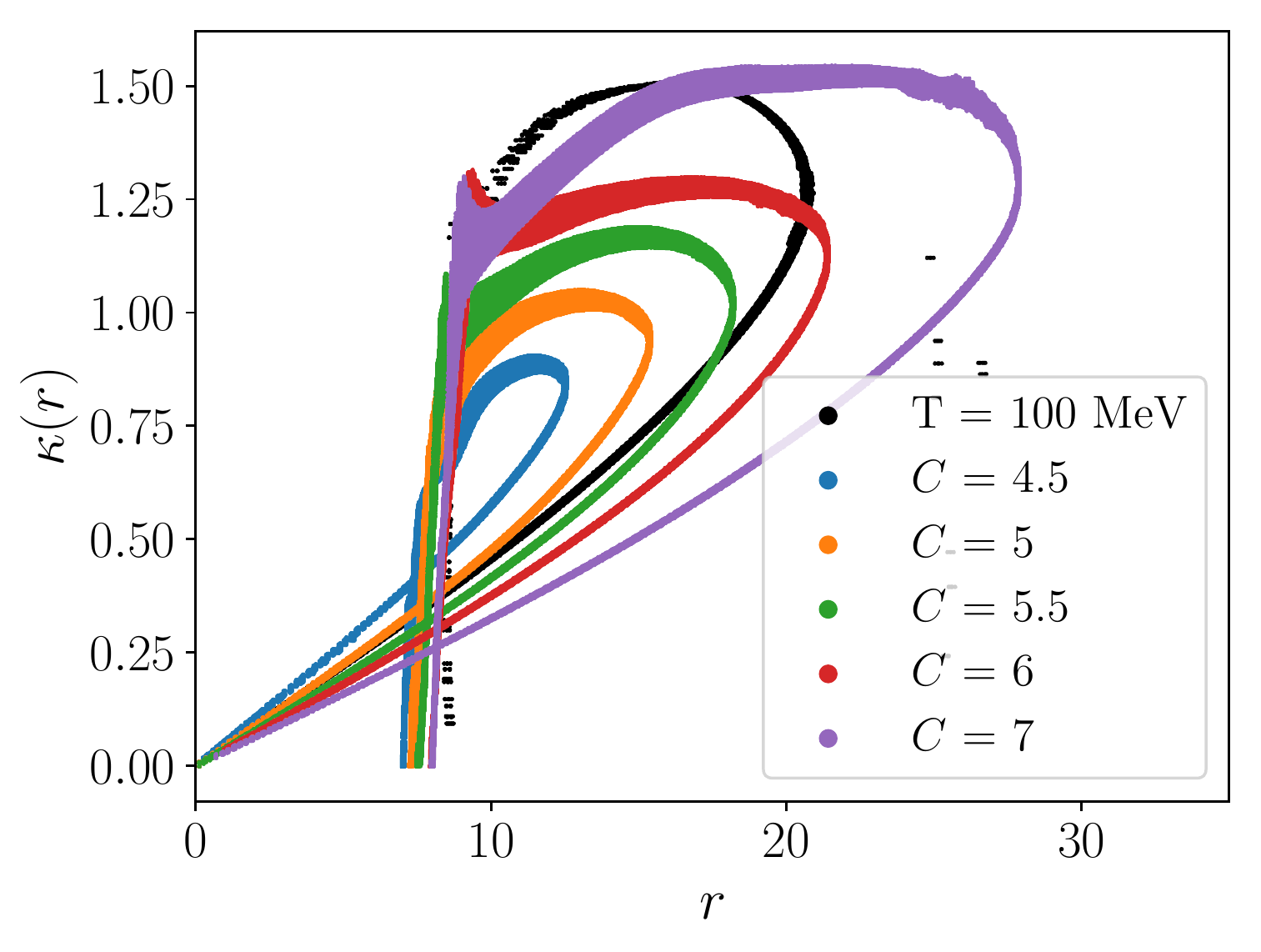}}
\caption{(a) The isotherms, marked by the dotted lines with the corresponding temperature in MeV,  compared with the
improved freezeout surfaces corresponding to the condition (\ref{eqn_FO_condition}), with the values of the parameter $C$  indicated on the insert. (b) shows the
transverse rapidity $\kappa(r)$ on the $T=100\, \text{MeV}$  isotherm (black) with those corresponding to improved FOS. 
}
\label{fig_FOS}
\end{center}
\end{figure}

In the upper plot of Fig. \ref{fig_FOS}
we compare the isotherms with the surfaces obtained from the  freezeout condition (Eq. \ref{eqn_FO_condition}). While having qualitatively similar shape, they generally correspond
to smaller $r$ and larger $\tau$. 
 The corresponding
distribution of the transverse flow rapidity on these surfaces is shown in 
the lower part of the figure.
 One can see that improvement of the freezeout condition leads to quite
significant 
enlargements of the  ``outer edge" plateaus. For example, the curve with $C=7$ provides the
transverse rapidity $\kappa\approx 1.4$ 
(needed for description of the high $p_\perp$ tail) 
for larger range of distances, from $r=8\,\, \text{fm} $ to   $r=27\,\, \text{fm} $. 

Let us emphasize again, that the improved FOS's depend on the cross section
for the particular species. For example, the difference between  those for a nucleon and $\phi$ meson is more than an order of magnitude, and even in the power $1/P$ they can be as large as 50\% difference of $C_i^{1/P_i}$, roughly corresponding to the range of parameters shown. If so, in spite of similar mass, the spectra of $p$ and $\phi$ at large $p_\perp$ 
should be very different, with the ``outer edge" component nearly absent in the latter case. 

%\section{Spectra at large $p_\perp$} 

\section{Extension to non-central collisions}
\begin{figure}[htbp]
\begin{center}
\includegraphics[width=7cm]{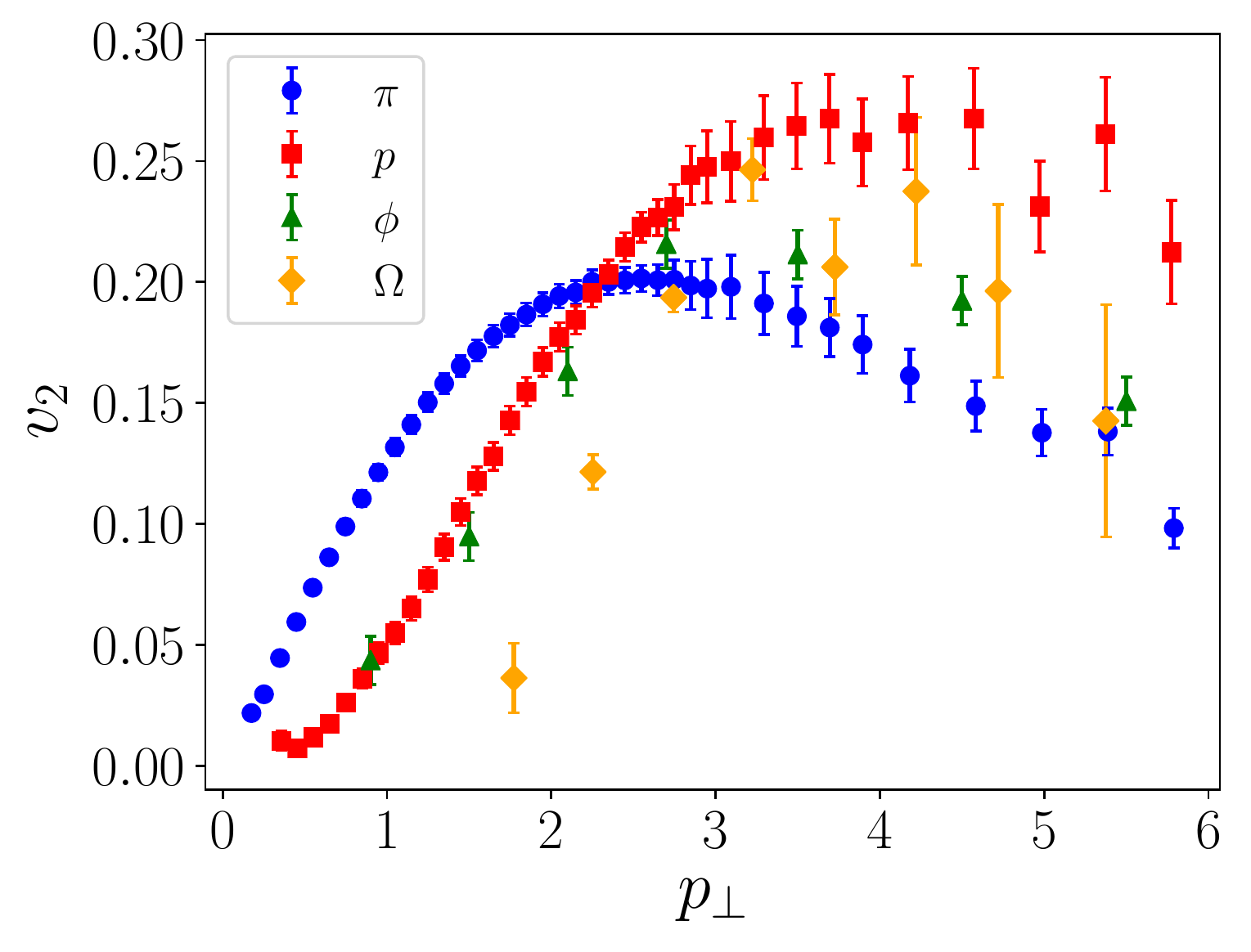}
\caption{Elliptic flow parameter $v_2$ for $\pi,p,\phi,\Omega$
identified hadrons, shown by blue circles, red squares, green triangles and orange diamonds,
respectively. The data are 
 from ALICE collaboration 
 \cite{1405.4632}, for the centrality bin 20-30\%.}
\label{fig_v2_alice}
\end{center}
\end{figure}

Let us start with the elliptic flow data, for various secondaries, in the same range of transverse
momenta as discussed in the previous chapters.  In  Fig. \ref{fig_v2_alice}, we have compiled   the ALICE data for $v_2(p_\perp)$ for some secondaries. The general trend is that $v_2$ {\em decreases} at $p_\perp>3 \, \text{GeV}$, after it reaches
a maximum.  Before going into details, let us outline a proposed explanation of this trend:
the azimuthal asymmetry of the flow is large in the ``lid" region, but is very small at
the ``outer edge". 

We also would like to argue that the ``outer edge" FOS and thus flow magnitude
is different for different secondaries, while their are much more similar at the ``lid". 
For this reason we included $v_2$ for  $p$ and $\phi$:
they are nearly identical below the maximum, at $p_\perp\approx 2.5\, \text{GeV}$, dominated by the ``lid" region, but  are different above it. We remind that these particles have similar mass
but very different collision rates with (mostly pion) background matter. 

%We expect that this observed downward trend of $v_2$ is due to the fact that the ``outer edge" part of the
%FOS has smaller angular asymmetry, as compared to the ``lid" part. 

To study whether the proposed explanation can be justified from standard hydrodynamical
approach,
 we  generated hydro output corresponding to non-central collisions. We use
 the same
centrality bin 20-30\% as the data just discussed. For this hydrodynamical solution, we  constructed 
both types of the FOS --the
isotherms as well as those
corresponding to the freezeout condition (Eq. \ref{eqn_FO_condition}) -- and calculated
the flow patterns on these surfaces.

\begin{figure}[h!]
\begin{center}
\subfigure[]{\includegraphics[width=.5\textwidth]{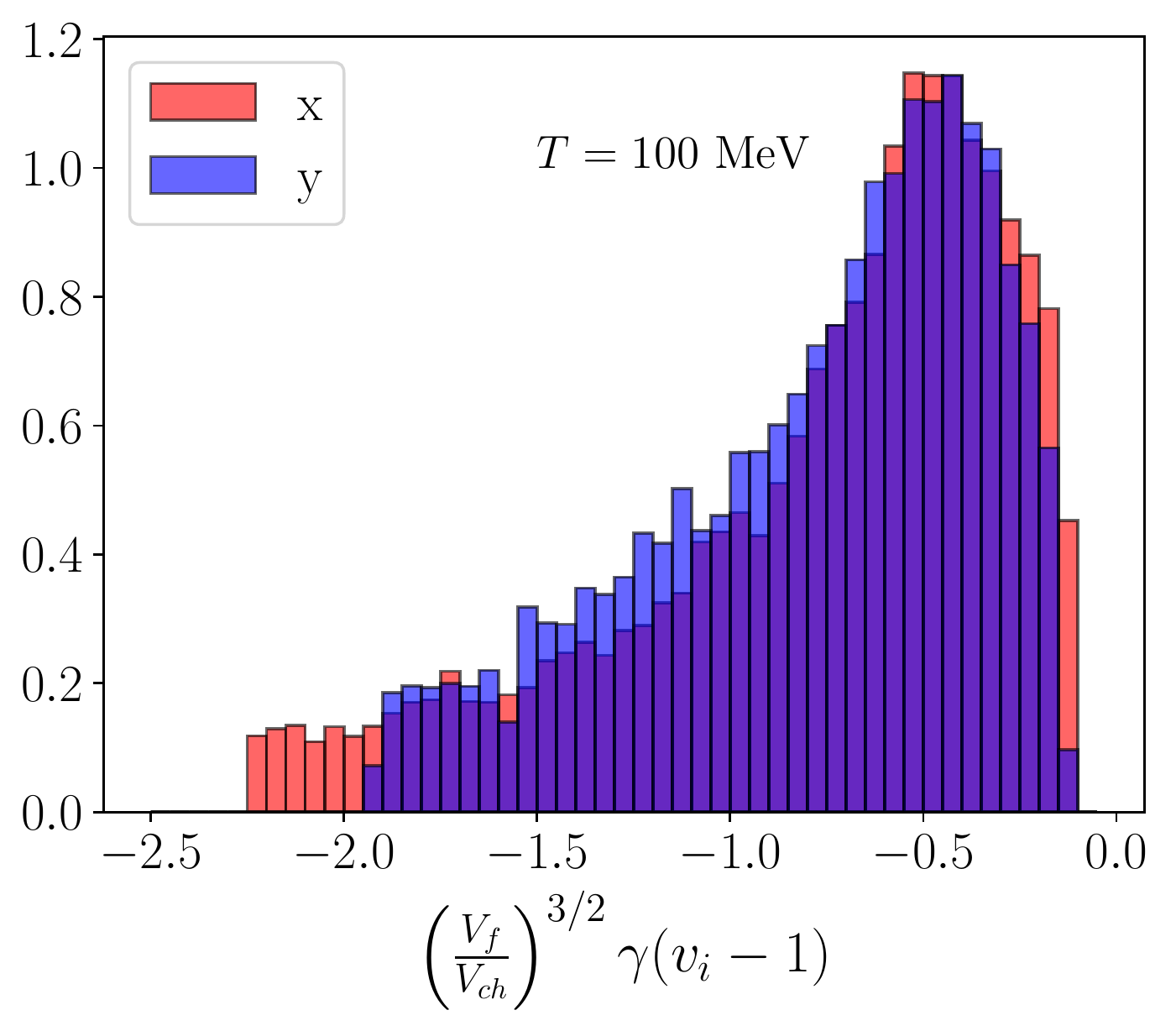}}
\subfigure[]{\includegraphics[width=.5\textwidth]{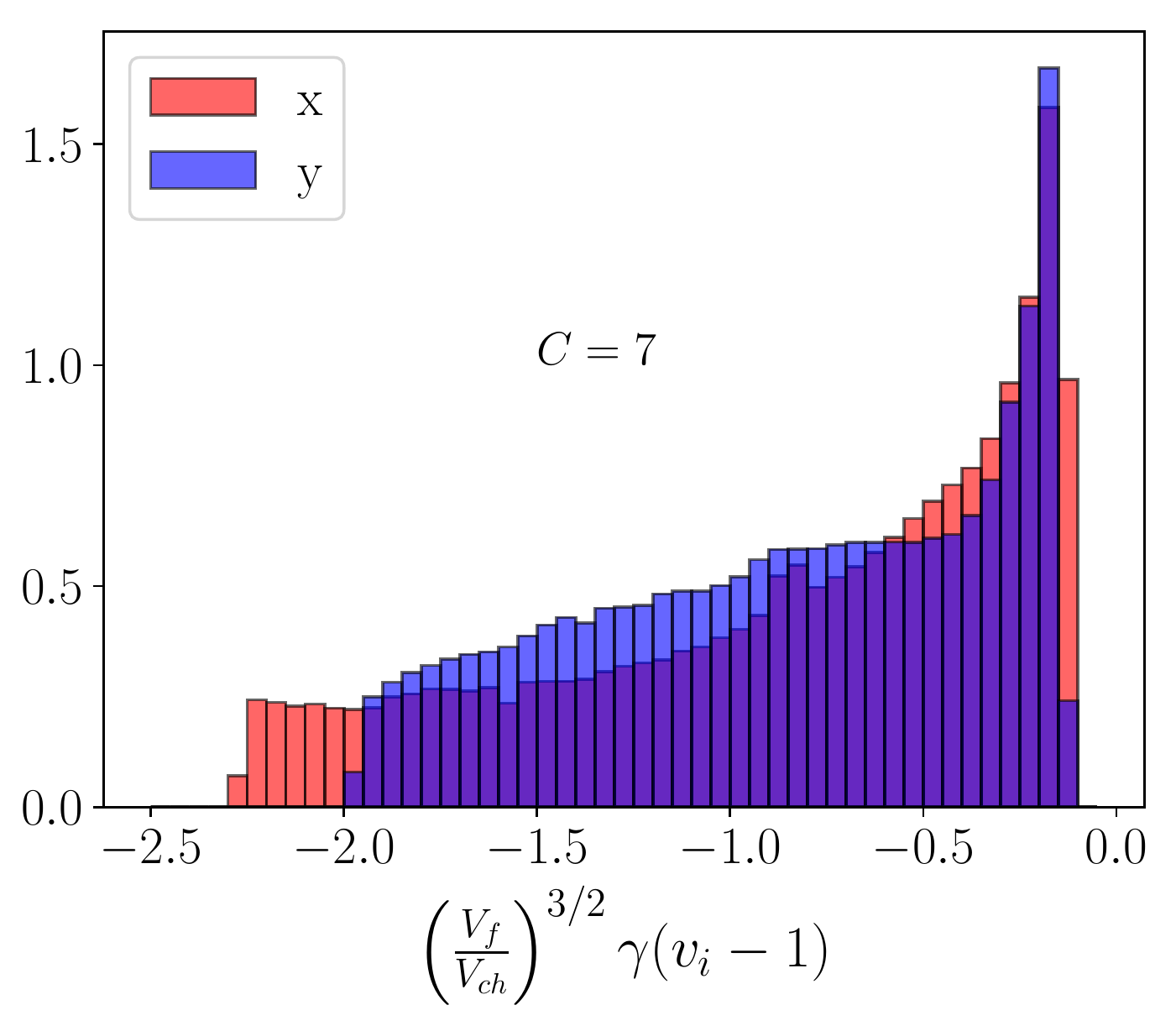}}
\caption{The distribution over $u_0-u_x$ (dark blue) and $u_0-u_y$ (light red)
on the FOS with (a) $T=100\, \text{MeV}$  and (b) the improved surface corresponding to 
condition Eq. (\ref{eqn_FO_condition}) with $(\partial_\mu u^\mu)^{1/3.5}=7T$.
}
\label{fig_xy}
\end{center}
\end{figure}

In Fig. \ref{fig_xy} we show the distributions of the flow rapidity in $x$ and $y$ directions, for both
types of surfaces. Specifically, to focus on azimuthal asymmetry, we show the
 distributions over 
$ u_0+u_x$ and $ u_0+u_y$ at the isotherm $T=100 \, \text{MeV}$ and the improved FOS with $(\partial_\mu u^\mu)^{1/3.5}=7T$. The choice of the quantity is due to the fact that for very ultrarelativistic secondaries, the Boltzmann factor $\exp(-p^\mu u_\mu/T)$ can be simplified to
 $\exp\big( (p/T)( u_0+u_x)\big)$ when momentum is in the $x$
direction, and similarly for $y$ direction. The comparison shows that: \begin{enumerate}[(i)]
\item the  improved FOS
has more pronounced  peaks at the r.h.s. of the plot, the region most important for
spectra at large $p/T$; 
\item the asymmetry
between $x$ (in the direction of the impact parameter) and $y$ distributions  for the improved FOS is
significantly reduced.
\end{enumerate}

While both these trends  are in the direction supporting the proposed explanation of $v_2(p_\perp)$ behavior, 
 unfortunately quantitatively it still does not work: 
  even the reduced asymmetry at the  improved FOS still results in
the values of the elliptic flow parameter $v_2(p_\perp>2.5\, \text{GeV})\approx 0.3-0.4$, which is larger than observed. We also do not observe the marked decrease of $v_2$ at high 
$p_\perp$ observed in the data.
We thus conclude that,  to better understand particle distributions at
the ``outer edge" of the fireball, more work is needed.

\section{Summary and discussion}
This paper was motivated by two observations. First, the thermal distribution
over particle masses correctly describe the data for many orders of magnitude,
at least up to the $M=4\, \text{GeV}$ of $^4$He, and maybe even further.
 One may also think that thermal exponential distribution in mass 
implies also thermal exponential distribution in
{\em energy}, in frames co-moving with matter. If so, the experimental spectra can be expressed  as
 a convolution of near-thermal spectrum with appropriate distribution over collective
 flow velocities. 
 
 Our second motivation, coming from spectra themselves, is that
 their transition to power-like behavior only at $p_\perp=5-6\, \text{GeV}$, and 
 below that we see exponential with 
  very smooth growth of ``effective temperature." Its explanation via ``blue shift" 
  suggest magnitude of flow compatible with values reached at the ``outer edge" of the fireball.
 Phenomenologically, a simple ``lid" plus ``outer edge" model does well, reproducing
spectra of secondaries with different mass, such as pions and nucleons.

 To understand the phenomenon further, we then turn to known analytic solutions. Dismissing 
  Gubser's  solution due to its unphysical behavior at the ``outer edge,"   
we then show that the ``rarefaction fan" solution, although 1+1 dimensional, can be used
to understand the edge of numerical solutions to hydrodynamics.  In particular, a 
nontrivial feature is that the isotherms and the
constant flow rapidity lines are  nearly identical.

The next step we took is to use 
an improved freezeout condition, including not only temperature but also
 the matter expansion rate defined via  $\partial_\mu u^\mu$. The corresponding modifications of
the FOS are studied, and the flow of magnitude needed to explain the  spectra
in central collisions is indeed observed on these surfaces. 
 
 Turning to non-central collisions we suggest that the observed decrease of $v_2$ at $p_\perp > 2.5\, \text{GeV}$ is due to very small azimuthal asymmetry of the flow at the ``outer edge".
 We found that the improved FOS does indeed have smaller asymmetry than isotherms,
 but it was still too large to explain the data. 
 
 Apparently, more work is needed to understand the conditions at the  ``outer edge"
 of the fireball. 
 
 Standard hydro-based models use cascade ``afterburners," but since we now discuss
 tails of the spectra with probabilities down to something like $10^{-6}$, this approach is not statistically feasible.  

The collision rate corresponding to FOS we use implies  
large mean free paths of particles at kinetic freezeout  about 
$$ \tau_\text{coll}\sim l_\text{m.f.p.}\sim 
7\, \text{fm}$$
While it may appear large, note that at that stage the fireball is still much larger than this value, with
time reaches 30 fm/c and radius about 20 fm.  So one does not need to do a cascade
of the whole system; rather, a cell of this size should be enough.

We would like to suggest relatively simple modifications of standard hydro may be significant increase of viscosities, bulk and shear, at the late stages. Another can be
 a substitution of FOS by its  ``coarse grained" version, on the scale defined by $l_\text{m.f.p}$. 
 
 Further improvement can involve anisotropic distribution in a cell,  in the
 frame co-moving with the flow. It is clear that isotropic thermal distribution
 should become anisotropic, given the temperature gradient. 
 Another effect indicated by  Ref. \cite{Teaney:2013gca} is that not only
  the scalar expansion rate 
$\partial_\mu u^\mu$ we used above needs to be included, but the whole tensor of gradients $\partial_\mu u^\nu$ as well.  So far,
 deformation of the thermal distribution due to it has been only studied when the deformation is small. 

\appendix

\section{General expressions for rapidity-independent and axially symmetric flows}

Our (relatively standard) assumption is  the (longitudinal) {\em rapidity independence} of flow,
and that it is simply equal to spatial rapidity, a la Bjorken. For this case
one can integrate the longitudinal extent of the fireball \cite{Heinz:2004qz}.
\be   {dN \over dy dp_\perp^2 d\phi}={2g_i \over (2\pi)^3}  \int d^2 r_\perp \tau_f(r_\perp) e^{\mu/T}
   e^{(\vec u_\perp \vec p_\perp)/T}  \label{eqn_bessel1}
   \ee
$$ \times   \big[  m_\perp K_1(\beta_\perp)-(\vec p_\perp \vec{\nabla}_\perp \tau_f) K_0(\beta_\perp)
   \big]\,,
$$
   where $m_\perp^2=p_\perp^2+m^2$. Since in this work we focus
   on large momenta, we assume Boltzmann statistics with a single exponent,
and   the arguments of the Bessel functions $K_1,K_0$ are the temporal part $\beta_\perp=m_\perp u_0/T$ of the product
of   the 4-velocity of the flow $$u_\mu={1 \over \sqrt{1-v^2}}(-1,\vec v)$$ to 4-momentum. Note that the temperature $T$ is not written as a function of the position, since the freezeout surface
is usually approximated by the isotherm with constant $T=T_f$.

Further simplification is possible when collisions are assumed to be exactly central, with zero
impact parameter $b=0$, since the cup is round and analytic integration over  azimuthal angle $\phi$
   can be done. The result \cite{Heinz:2004qz} includes the second set of Bessel functions 
   \be 
  {dN \over dy dp_\perp^2 }={g_i \over \pi^2} \int d r_\perp r_\perp \tau_f(r_\perp)e^{\mu/T}
 \label{eqn_bessel2}   \ee
  $$   \times   \big[  m_\perp K_1(\beta_\perp)I_0(\alpha_\perp)-p_\perp \left({\partial \tau_f 
  \over \partial r_\perp}\right) K_0(\beta_\perp)I_1(\alpha_\perp)   \big] $$
   with the argument being the 3-d part of the $u_\mu p^\mu$ product, namely $\alpha_\perp=(\vec u \vec p_\perp)/T)$.
    
%%%%%%%%%%%%%
%
%\section{The  local anisotropy to the first order in gradients}
%The viscous corrections to the stress tensor to the first order in gradients
%is known as the Navier-Stokes term, which in relativistic hydrodynamics takes the form
%\be \delta T_{\mu\nu}^{(1)}=-\eta \sigma_{\mu\nu} \ee
%with $\eta$ being the shear viscosity. 
%
%Definition of the sigma tensor is using the following standard notations
%for the projector to the flow restframe and projector to symmetric traceless tensors
%in this frame
%\be  \Delta^{\mu\nu}=g^{\mu\nu}+ u^\mu u^\nu \ee
%\be \langle A^{\mu\nu} \rangle={1 \over 2}    \Delta^{\mu}_\rho  \Delta^{\nu}_\sigma (A^{\rho\sigma}+A^{\sigma\rho})-{1 \over d-1}  \Delta^{\mu\nu} \Delta_{\rho\sigma }A^{\rho\sigma}
%\ee
%namely
%\be \sigma_{\mu\nu}=2  \langle u_{\mu ;\nu} \rangle \ee
%with semicolon, as usual, standing for covariant derivative. 
%
%The local anisotropy follows from the kinetic equation 
%\be p^\mu \partial_\mu f_p=- {E_p\over \tau_R}\big( f_p -f_{(0)}\big) \ee

\section{Implementing nonzero chemical potentials after chemical freezouts}
Ignoring small presence of baryon number at mid-rapidity, one usually assume
that before chemical freezeout ($T_\text{ch}\approx 156\, \text{MeV}$) all species have zero chemical potential. 

However 
 after that, the inelastic collisions are assumed to be absent and the particle numbers of all species  are assumed to
be conserved. If so, the thermal distributions are appended by  $nonzero$ chemical potential,
whose magnitude is calculated from particle number preservation. 
It is approximately given by the following relation
\be e^{\mu(T) \over T} ={V_\text{ch} \over V_T}\left( {T_\text{ch} \over T}\right)^{3/2}\,. \ee
Since it depends on $T$ only, it is a constant factor on isotherm FOS. But
  on the improved FOS's we use the temperature is no longer constant,
and thus this fugacity varies at different locations, so it therefore needs to be included in any averaging.

\section{Hadronic observables in our hydrodynamic calculations}

The hadronic observables produced in our hydrodynamic simulations for for 2.76 TeV Pb-Pb collisions with optical Glauber initial conditions and the experimental data are shown in Table \ref{hydro2760}. In general, we have good agreement with data. The $v_2$ of our calculations is smaller than that of experiment, which is a well-known result of using optical Glauber initial conditions.

\begin{table}[h!]
\centering
\caption{Comparison of calculated and experimental hadronic observables in 2.76 TeV Pb-Pb collisions, for hydrodynamics with bulk viscosity. Experimental data is the same as used in Refs. \cite{Ryu:2015vwa,Ryu:2017qzn}.}
\label{hydro2760}
\begin{tabular}{|l|l|l|}
\hline
                & hydro. calc.       & data         \\
                                \hline
$N_\text{pion}$                           & 309.1     & 307$\pm$20        \\
$\braket{p_T}_\text{pion}$& 0.508    & 0.512$\pm$0.017   \\
$v_2$                            & 0.0746 & 0.0831$\pm$0.0034 \\ 
\hline
\end{tabular}
\end{table}

{\bf Acknowledgements} This work was supported by the U.S. Department of Energy under Contract No.
DE-FG-88ER40388. We acknowledge help by J.-F. Paquet, who kindly guided us in the use of 
hydrodynamical numerical solutions generated using the MUSIC code.

\end{document}